\documentstyle[prd,aps,psfig]{revtex}
\begin{document}
\draft
\input epsf
\twocolumn[\hsize\textwidth\columnwidth\hsize\csname
@twocolumnfalse\endcsname
\title{{\hfill \small DSF-4-2001, MPI-PhT/2001-3, hep-ph/0102080}\\ $~$\\
Bose--Einstein condensation at reheating}
\author{ G. Mangano$^{(1)}$, G. Miele$^{(1)}$, S. Pastor$^{(2)}$, and
M. Peloso$^{(3)}$}

\address{$^{(1)}${\it Dipartimento di Scienze Fisiche, Universit\`{a} di Napoli
{Federico II} and INFN, Sezione di Napoli, \\ Complesso Universitario di
Monte S. Angelo, Via Cintia, I-80126, Naples, Italy.}}
\address{$^{(2)}${\it Max-Planck-Institut f\"{u}r Physik} (Werner-Heisenberg-Institut),
F\"{o}hringer Ring 6, D-80805, Munich, Germany.}
\address{$^{(3)}${\it Physikalisches Institut, Universit{\"{a}}t Bonn Nussallee 12,
D-53115 Bonn, Germany.}}

\maketitle
\begin{abstract}
We discuss the possibility that a perturbative reheating stage after
inflation produces a scalar particle gas in a Bose condensate state,
emphasizing the possible cosmological role of this phenomenon for symmetry
restoration.
\end{abstract}
\pacs{PACS number: 98.80.Cq}
\vskip2pc]
\def\gsim{\;\raise0.3ex\hbox{$>$\kern-0.75em\raise-1.1ex\hbox{$\sim$}}\;}
\def\lsim{\;\raise0.3ex\hbox{$<$\kern-0.75em\raise-1.1ex\hbox{$\sim$}}\;}

\section{Introduction}

At the end of inflation, the Universe was in a very cold, low entropy
phase. In particular, its energy density was still dominated by the vacuum
energy of the fields which were previously driving inflation. The process
of converting this energy into a hot thermal bath of matter and radiation
is known as {\em reheating}, and it represents one of the most active areas
of research in inflationary cosmology. In the last ten years, following the
works \cite{preh1,preh2}, it has been shown that the first stages of
reheating can be characterized by interesting nonperturbative phenomena,
according to the particular theory one is considering. However, all these
different possibilities share the common feature that they are followed by
a more standard stage of perturbative particle production leading to the
formation of the thermal bath. In the present work we focus on the
possibility that after inflation some light boson field $\chi$ may
experience a Bose-Einstein condensation, after production of the
corresponding quanta via a perturbative reheating stage. This situation may
be naturally realized if these quanta carry a charge which is either
exactly conserved or weakly violated by interaction processes with all
species produced at reheating. Condensation may then occur if a
sufficiently large value for this charge is released during reheating .

The first condition is eventually a statement on the lagrangian density
governing the $\chi$ field dynamics. As an example we consider a simple
model of a complex scalar field $\chi$ whose interaction term is of the
form of a quartic self-interaction monomial, invariant under a global
$U(1)$ symmetry. Despite of its simplicity we look at this model as a
prototype of much more complicated and realistic theories, as for example
supersymmetric extensions of the electroweak standard model, which admit
scalars as elementary excitations and conserved, or weakly violated,
charges.

Specific mechanisms for production of a non zero and sufficiently large
value for a definite charge at reheating have been widely discussed in the
literature, aimed to drive, for example, efficient ways to produce baryon
and lepton numbers. Among these, the Affleck-Dine scenario \cite{ad} has
been shown to be able to produce very large baryon and lepton asymmetries.
In this case scalar particles can naturally develop a Bose condensate.

In view of these considerations, we think that it is an interesting issue
to study under which conditions a Bose condensate may actually form.
Section II and III are devoted to this subject. In particular, in section
II we report a fully general analysis of conditions necessary for the
formation of the condensate, while in section III we discuss its actual
formation in a cosmological context. We finally consider in section IV the
possibility that the condensate may have an interesting cosmological
implication, and leads to the phenomenon of symmetry restoration. This may
reintroduce a monopole problem, since when symmetry eventually breaks
again, once the expansion of the Universe dilutes the condensate,
topological defects can be formed.

\section{Conditions for a condensate}

In this section we discuss the conditions leading to the formation of a
condensate. We consider a distribution of quanta of a light complex scalar
field $\chi\,$, whose mass $m_\chi$ is assumed to be dominated by radiative
effects. We also assume that the initial distribution has a
particle/antiparticle asymmetry, $Q \equiv n_\chi - n_{\bar \chi}$, and
that the charge per comoving volume is conserved (or only very weakly
violated) by the self-interactions among these quanta.~\footnote{We neglect
here interactions with other fields; see the next section for a more
general discussion of the system in a cosmological setup.} Without loss of
generality, we consider $Q > 0\,$. The production of this charge will be
discussed in the next section.

The evolution of the system is of course strictly dependent on these
self-interactions, since they determine the thermalization time-scale as
well as the radiative mass. As an example, let us consider the
interaction term
\begin{equation}
{\cal L}_{\rm int} = - \, h \, \vert \chi \vert^4 \;\;\;.
\label{inter}
\end{equation}
At lowest order in $h$, this term is  responsible for $2 \leftrightarrow 2$
scatterings. Combining more vertices, processes which change the total
number of particles $n=n_\chi + n_{\bar \chi}\,$ are possible, preserving
however the charge $Q$. The efficiency of these processes in establishing a
kinetic equilibrium, i.e. the thermalization time-scale, will be discussed
in the next section. Here we consider the asymptotic equilibrium state in
presence of a {\it fast} thermalization.

We start studying the case where a condensate does not form. The
equilibrium distribution functions are therefore given by
\begin{equation}
f_{\chi \,,\, {\bar \chi}}^{BE} \left( {\bf p} \right) = \left[  {\rm
e}^{\, \beta \left( E \mp \mu \right)} -1
\right]^{-\,1} \;\;\;.
\label{bosein}
\end{equation}
In this expression, $E \equiv \sqrt{{\bf p}^2 + m_\chi^2(T)}$ denotes the
energy of the quanta, while $\beta = 1/T$ is the inverse
temperature~\footnote{We use natural units $c=\hbar=k_B=1$.}. The presence
of the chemical potential $\mu$ is related to a nonvanishing charge $Q\,$.
If $Q$ and $\mu$ are both positive, the minus sign in Eq.~(\ref{bosein})
refers to particles $\chi\,$, and the plus sign to antiparticles ${\bar
\chi}\,$. The equality $\mu_\chi = - \, \mu_{\bar \chi} \equiv \mu$ holds
in presence of effective number violating interactions such as $\chi \,
{\bar
\chi} \, \chi \, {\bar \chi} \leftrightarrow \chi \, {\bar \chi}\,$.

While the distribution function $f_{\bar \chi}^{BE}$ is always regular, we
must impose $\mu < m_\chi\,$ to ensure the regularity of $f_\chi^{BE}\,$.
The values of $\beta$ and $\mu$ can be determined by the two covariantly
conserved quantities of the system, that is the total energy $\rho$ and
charge $Q$ densities
\begin{eqnarray}
\rho &=& \int  \frac{d^3 {\bf p}}{\left( 2 \,
\pi \right)^3}  \; E \; \left[ f_\chi^{BE} \left( {\bf p} \right)+
f_{\bar \chi}^{BE} \left( {\bf p} \right) \right] \;\;\;, \label{rho}
\\
Q &=& \int \frac{d^3 {\bf p}}{\left( 2 \,
\pi \right)^3} \; \left[ f_\chi^{BE}\left( {\bf p} \right) -
f_{\bar \chi}^{BE}\left( {\bf p} \right)
\right]
\;\;\;.
\label{cond}
\end{eqnarray}

Let us consider the ratio
\begin{equation}
R \left( \mu/T \,,\, m_\chi / T \right) \equiv \frac{Q}{\rho^{3/4}} \;\;\;.
\label{ratio}
\end{equation}
Notice that since we deal with radiation this ratio is unaffected by the
universe expansion, since both quantities $Q$ and $\rho^{3/4}$ scale as the
inverse third power of the scale factor, so that their ratio $R$ is
constant. It is easy to see that $R$ monotonically increases with $\mu/T$.
In particular, it ranges between $0$ and the critical value $R_{cr} \left(
m_\chi/T
\right)$ $\equiv R\left(m_\chi
/T,m_\chi /T
\right)$ as $\mu$ ranges between $0$ and $m_\chi\,$. We have thus a simple
criterion to understand whether a condensate forms. From the initial
conditions for the system at reheating we calculate the quantity
$R=Q/\rho^{3/4}\,$, which is conserved by the interactions among the quanta
$\chi$ and ${\bar
\chi}\,$. If this value is below $R_{cr} \left( m_\chi /T \right)$,
the equilibrium distributions are just of the form~(\ref{bosein}), that is
the condensate does not form. When instead the initial conditions are such
that $R$ is greater than $R_{cr}\left( m_\chi /T \right)$,
Eqs.~(\ref{bosein}) cannot represent the final distributions, but an
additional singular term is present in $f_\chi({\bf p})$
\begin{equation}
f_\chi \left( {\bf p} \right) = f^{BE} \left( {\bf p} \right) + (2
\pi)^3 Q_c \, \delta^3 \left( {\bf p} \right) \;\;.
\label{disttot}
\end{equation}
This shows the appearance of a condensate at zero momentum. The physical
interpretation of this is very simple. Loosely speaking, $R > R_{cr} \left(
m_\chi /T \right)$ corresponds to a case in which a too large charge $Q$ is
present to be stored in a regular distribution function with the given
energy density $\rho\,$. The exceeding part of this charge, denoted by
$Q_c$, gives rise to the condensate.

\begin{figure}
\centerline{\psfig{file=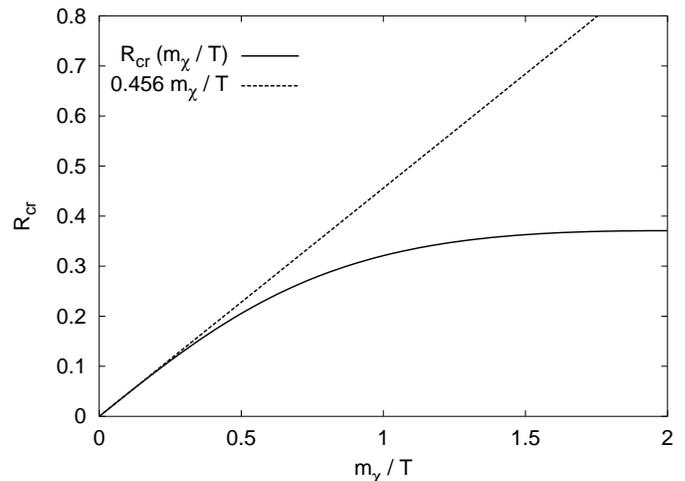,angle=-90,width=.52\textwidth}}
\caption{The value of $R_{cr}\left( m_\chi /T \right)$ versus $m_\chi / T\,$
both numerically (solid line)  and with an analytic approximation valid at
small arguments (dotted line). Initial conditions such that $R >
R_{cr}\left( m_\chi /T \right)$ indicate the formation of a condensate.}
\label{fig1}
\end{figure}

The value of $R_{cr} \left( m_\chi / T \right)$ can be obtained from
expressions (\ref{rho}) and (\ref{cond}) in the limit $\mu \rightarrow
m_\chi$ for the distribution (\ref{bosein}). The calculation can be
performed numerically, or analytically when $m_\chi / T $ is much smaller
than one. In this regime we find
\begin{eqnarray}
\rho &=& T^4 \left( \frac{\pi^2}{15} + {\cal O} \left( \frac{m_\chi^2}
{T^2} \right) \right)~~~, \label{approxq} \\ Q &=& T^3 \left( \frac{1}{3}
\, \frac{m_\chi}{T} + {\cal O} \left(
\frac{m_\chi^3}{T^3}
\right) \right) ~~~, \label{approxrho}
\end{eqnarray}
such that the scaling $R_{cr}\left( m_\chi /T \right) \simeq 0.456 \,
m_\chi / T\,$ is expected for $m_\chi /T \ll 1\,$. We plot in
Figure~\ref{fig1} both the exact numerical behavior and the analytical
approximation for $R_{cr}\left( m_\chi /T \right)$. As we see, the
analytical approximation is very good even for $m_\chi$ as large as $0.1
\div 0.2 \, T\,$.

If the field $\chi$ is very light, its mass will be dominated by radiative
effects. In particular, from the interaction~(\ref{inter}) among particles
in the thermal distributions~(\ref{bosein}) we can perturbatively evaluate
the thermal mass for $h \ll 1$. From one loop correction to $\chi$ propagator
one easily find
\begin{eqnarray}
m_\chi^2 &=& 4 h \int \frac{d^3 \bf{p}}{(2 \pi)^3 2 E}
\left( f^{BE}_\chi \left( {\bf{p}} \right) + f^{BE}_{\bar{\chi}}
\left( {\bf{p}} \right) \right)\;\;\;,
\end{eqnarray}
and hence
\begin{equation}
\frac{m_\chi \left( T \right)}{T} \sim \sqrt{\frac{h}{3}} \;\;\;,
\label{masschi0}
\end{equation}
Therefore we see that a condensate forms whenever the initial conditions
result in
\begin{equation}
R \gsim \, 0.26 ~h^{1/2} \;\;\;.
\label{cond1}
\end{equation}

In the next sections we will discuss scalar condensation in a cosmological
context. It is important to stress again that the expansion of the Universe
does not modify the above condition~(\ref{cond1}), at least until the
temperature drops below the value of the (non-thermal) mass of $\chi$.

We now consider the case $R> R_{cr}\left( m_\chi /T \right)$, so that a
condensate forms. In general with the distribution function for the scalars
$\chi$ given by Eq. (\ref{disttot}), both charge $Q$ and energy density
$\rho$ are shared by the regular part $f^{BE}_{\chi,\bar{\chi}}$ and the
condensate. The analysis greatly simplifies in the limit where most of the
the charge $Q$ is stored in the condensate. It is worth noticing that this
is also the case we are more interested in. When a large charge is
initially produced, the formation of the Bose condensate may lead to
cosmologically relevant effects. In this case, we have $n_\chi \sim Q_c \gg
n_{\bar{\chi}}$ and $Q_c \gg Q_{th}$, where we have defined with $Q_c$ and
$Q_{th}=Q-Q_c$ the charge accomodated in the condensate and in the thermal
distribution respectively. Whenever number density is dominated by the
condensate, evaluation of the thermal mass using Eq. (\ref{disttot}) gives
\begin{eqnarray}
&m_\chi^2 &\sim 4 h \int \frac{d^3 \bf{p}}{(2 \pi)^3 2 E}
\left( f^{BE}_{\chi} \left( {\bf{p}} \right)  + f^{BE}_{\bar{\chi}}
\left( {\bf{p}} \right) \right. \nonumber \\
 &+& \left. (2
\pi)^3 Q_c \, \delta^3 \left( {\bf p} \right) \right)
 \sim  2 h \frac{Q_c}{m_\chi} \;\;\;,
\end{eqnarray}
and therefore
\begin{equation}
m_\chi \left( Q_c \right) \sim \left( 2 h \, Q_c
\right)^{1/3}.
\label{masschi}
\end{equation}
This result of course holds when $Q_c^{1/3} > T$. More generally, for
smaller $Q_c^{1/3}/T$ it is easy to check that a very good approximation
for the thermal mass is given by
\begin{equation}
m_\chi \sim \left( 2 h Q_c + \frac{1}{3} h T^3\right)^{1/3} \;\;\;.
\label{truemass}
\end{equation}
The total number density can be estimated as follows
\begin{equation}
n \sim Q_c + n_{th}(Q_c,T) \;\;\;,
\label{number}
\end{equation}
where the thermal contribution $n_{th}(Q_c,T)$ can be numerically evaluated
using the expression $f^{BE}_{\chi,\bar{\chi}} \left( {\bf{p}} \right)$
with $\mu_\chi=m_\chi$, and $m_\chi$ given by Eq. (\ref{truemass}). In
particular when the temperature is larger than the effective mass, though
it is still smaller than $Q_c^{1/3}$
\begin{equation}
(2 h Q_c)^{1/3}<T<Q_c^{1/3} \;\;\;,
\label{condmass}
\end{equation}
we have
\begin{equation}
n_{th} \sim \frac{2 \zeta(3)}{\pi^2}T^3 \;\;\;.
\label{number0}
\end{equation}
Similarly the energy density can be written as
\begin{equation}
\rho \sim m_\chi Q_c + \rho_{th} (Q_c,T) \;\;\;.
\label{rhotot}
\end{equation}
where $\rho_{th}(Q_c,T)$ and $m_\chi Q_c$ are the thermal contribution and
the energy density stored in the condensate respectively.

\begin{figure}
\centerline{\psfig{file=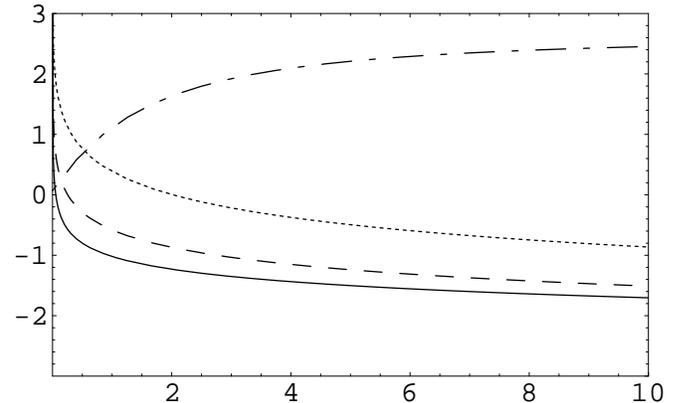,angle=0,width=.5\textwidth}}
\caption{The ratios $Q_{th}/Q_c$ (solid line), $n_{th}/Q_c$ (dashed line),
$\rho_{th}/m_\chi Q_c$ (dotted line) and $R$ (dash-dotted line), in
logarithmic scale, versus $Q_c/T^3$. $Q_{th}$ is the charge stored in the
thermal distribution. Results are shown for $h=0.01$.}
\label{termici}
\end{figure}

In Figure 2 we plot the values for the ratios $Q_{th}/Q_c$, $n_{th}/Q_c$,
$\rho_{th}/m_\chi Q_c$, and $R$, as functions of the variable $Q_c/T^3$,
where $Q_{th}$ is the charge stored in the thermal component. Results
corresponds to $h=0.01$ and have been evaluated using Eq. (\ref{truemass}).
Notice that for $Q_c \sim T^3$ the energy density is still dominate by the
thermal distribution, while both the charge and number density receive
their main contribution by the condensate.

Changing the value of $h$ slightly affects all quantities, but it does not
change the monotonic decreasing behaviour as function of $Q_c/T^3$ shown in
Fig. 2. As an illustration we show in Fig. 3 how the above quantities
depends on $h$, at $Q_c/T^3=1$, choosing $h$ in the range $10^{-10} \div
10^{-1}$.

It is rather important to notice that the total energy density $\rho$
always redshifts as the one of radiation. While this is obvious for the
thermal part of the distribution, for the condensate it follows from the
dependence of $m_\chi$ on $Q_c$ and $T$ given in Eq.~(\ref{truemass}). As a
consequence, the energy density stored in the condensate evolves in time as
$Q_c^{4/3} \sim a^{-4}\,$, where $a$ is the scale factor of the Universe.
\begin{figure}
\centerline{\psfig{file=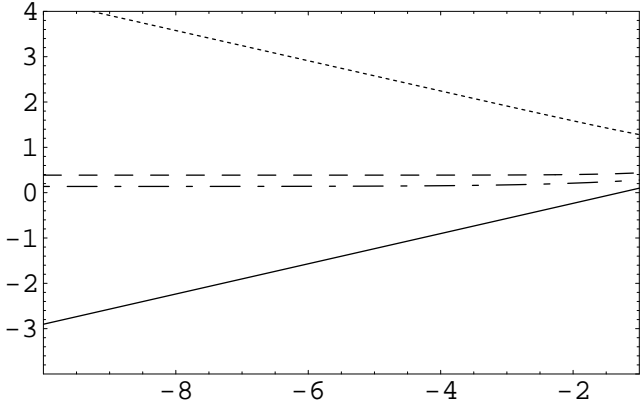,angle=0,width=.5\textwidth}}
\caption{The ratios $Q_{th}/Q_c$ (solid line), $n_{th}/Q_c$ (dashed line),
$\rho_{th}/m_\chi Q_c$ (dotted line) and $R$ (dash-dotted line), in
logarithmic scale, versus $log ~h$, at fixed $Q_c/T^3=1$.}
\label{termicih}
\end{figure}

As we have said, we will be mainly interested in the following in studying
the case where $Q_c$ represents the dominant contribution to $n$. This
condition is satisfied whenever $Q_c >n_{th}$. Using the behaviour of
$n_{th}$ and $\rho$ with $Q_c/T^3$ this condition can be cast in a lower
bound for the ratio $Q/\rho^{3/4}$
\begin{equation}
\frac{Q}{\rho^{3/4}} \gsim \frac{1}{3} \div \frac{1}{2}
\;\;\;,
\label{cond2}
\end{equation}
as $h$ varies in the range $10^{-10} \div 10^{-1}$. Condition (\ref{cond2})
is of course more severe than Eq. (\ref{cond1}), which merely states the
formation of a condensate regardless of its magnitude with respect to
$n_{th}$. The order of magnitude of this result, obtained by a numerical
study, can be also achieved analitically by noticing that for all values of
the coupling we have considered, when $n_{th} \sim Q_c$ the energy density
receives the dominant contribution by the thermal part of the $\chi
-\bar{\chi}$ distribution, so that
\begin{equation}
\rho \sim \rho_{th}= \frac{\pi^2}{15}T^4 \;\;\;.
\end{equation}
Using the fact that $\rho_{th}^{3/4}\sim 3 n_{th}$ (see Eq.
(\ref{number0})) we obtain that $Q_c$ dominates the number density when
$Q_c > n_{th} \sim \rho_{th}^{3/4}/3$, i.e. whenever $R \gsim 1/3$.

\section{Formation of the condensate}

In this section we discuss a possible scenario for the formation of a
condensate in the early stages of the Universe.~\footnote{For related
works, see~\cite{Semikoz1,Semikoz2,Semikoz3}.}. We start with a field
$\phi$ which is oscillating around the minimum of its potential
$V(\phi)$.~\footnote{We assume that the minimum of the potential
corresponds to $\phi=0$ and that anharmonic terms in the Taylor expansion
of $V(\phi)$ can be neglected, so that $\phi$ has a constant mass
$m_\phi\,$.} One interesting possibility is that $\phi$ is a modulus field
of a supersymmetric theory. SUSY models offer several candidates for moduli
fields, since their scalar potential almost unavoidably presents flat
directions if supersymmetry is preserved. Most of these directions are
expected to acquire a mass of the order of the supersymmetry  breaking
scale. Actually this scale, in the early Universe, can be very different
from the present one, due to the dynamics of the inflaton or of other
scalar fields during inflation and the first stages of reheating. See
\cite{gra1,gra2,gra3} for detailed discussions. Nevertheless, though
supersymmetry naturally offers a theoretical framework, we aim to discuss
the condensation phenomenon in a more general contest. For this reason we
leave the mass $m_\phi$ as a free parameter. Also for the initial amplitude
$\phi_0$ of the field we have some natural scales, as for example the
Planck mass, emerging in the context of supergravity models, or the
breaking scale of GUT theories. However, also for $\phi_0$ very different
values can be considered in principle, and we will also take it as an input
parameter.

At very early times the field $\phi$ is frozen at $\phi_0\,$, due to the
friction provided by the expansion of the Universe. When, at a time $t
\equiv t_0\,$, the expansion rate $H$ falls down $m_\phi\,$, the field
$\phi$ starts oscillating around the minimum $\phi=0\,$. The amplitude of
these oscillations decreases as $a^{-3/2}\,$. If this stage is already
radiation dominated, we then have
\begin{equation}
\phi \left( t \right) = \phi_0 \left( \frac{a_0}{a \left( t \right)}
\right)^{3/2} \;\;\;,\;\;\; H \left( t \right) = m_\phi \left( \frac{a_0}{a
\left( t \right)} \right)^2 \;\;\;,
\label{evol}
\end{equation}
where $a_0 \equiv a \left( t_0 \right)\,$.

If $\phi$ has a sufficiently long lifetime, its coherent oscillations will
dominate over the thermal background for $a > a_{dom} \equiv a_0 \left( 3 /
8 \pi \right) \left( M_{Pl}/\phi_0 \right)^2 \,$. After this time, the
energy density stored in the oscillations redshifts with a matter-like
behaviour, so that
\begin{equation}
H \left( t \right) = \sqrt{\frac{8 \pi}{3}} m_\phi \frac{\phi_0}{M_{Pl}}
\left( \frac{a_0}{a \left( t \right)} \right)^{3/2}
\;\;\;,\;\;\; a > a_{dom}
\;\;\;,
\label{hubble}
\end{equation}
while the amplitude $\phi \left( t \right)$ is still given by
Eq.~(\ref{evol}).

When eventually the $\phi$ field decays, it produces a reheating stage. In
particular, baryon  and lepton asymmetries can be generated in a rather
natural way through the Affleck-Dine mechanism~\cite{ad} if $\phi$ evolves
along a complex direction and has baryon and lepton violating couplings.
The same mechanism can be applied for the production of a generic charge
$Q\,$. Let us indeed suppose that the field $\phi$ mainly decays at the
time $t_d$ into the complex scalar $\chi$ considered in the previous
section. Through a mechanism  {\it \`{a} la} Affleck-Dine, these decays produce
an asymmetry
\begin{eqnarray}
Q &\equiv& n_\chi - n_{\bar \chi} = f \, n_\phi \left( t_d \right) \;\;\;,
\nonumber\\ n_\phi \left( t_d \right) &\equiv& \rho_\phi \left( t_d \right)
/ m_\phi = \phi_d^2 \, m_\phi \;\;\;.
\label{prochi}
\end{eqnarray}
Without entering into the details of the production, which is crucially
dependent on the specific coupling between $\phi$ and $\chi$ and on the
dynamics of the scalar field $\phi$, we have parametrized with $f$ in
Eq.~(\ref{prochi}) the charge produced per quantum of $\phi\,$. We have
also parametrized with $\phi_d$ the amplitude of the oscillations of $\phi$
at decay.

It is immediate to see how large $f$ should be in order to produce a
condensate in the $\chi$ quanta distribution. Since $Q/\rho^{3/4}
= f\,\rho_\phi \left( t_d
\right)^{1/4} / m_\phi\,$, we see that the necessary condition for a condensate,
Eq.~(\ref{cond1}), is met for
\begin{equation}
f^2 \gsim   0.07 \, h \, \frac{m_\phi}{\phi_d}
\;\;\;.
\label{cond3}
\end{equation}
The more stringent requirement that the number density of $\chi$ is dominated by
the condensate, Eq.~(\ref{cond2}), rewrites instead
\begin{equation}
f^2 \,  \gsim (0.1 \div 0.25)  \, \frac{m_\phi}{\phi_d} \;\;\;.
\label{qincond}
\end{equation}

\begin{figure}
\centerline{\psfig{file=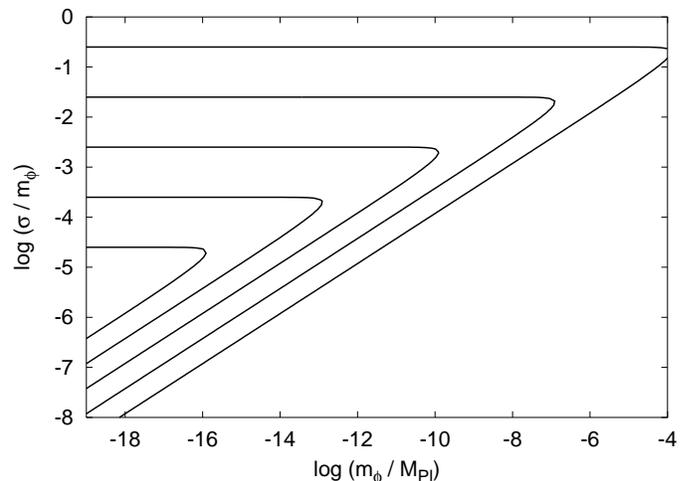,angle=-90,width=.52\textwidth}}
\caption{Contours for the ratio $\phi_d/m_\phi$ in the
$log(m_\phi/M_{Pl})$ - $log(\sigma/m_\phi$) plane. Continuous lines
correspond, from right to left, to $\phi_d/m_\phi=1,10,10^2,10^3$ and
$10^4$.}
\label{livelli}
\end{figure}

The interaction terms responsible for the decay of $\phi$ into $\chi$ will
also give a mass to $\chi-\bar{\chi}$ quanta through radiative corrections.
This mass vanishes once $\phi$ has decayed, so that the considerations of
the previous section (for example, Eq.~(\ref{truemass})) remain unchanged.
However, it could prevent $\phi$ from decaying, whenever $m_\chi \left(
\phi_d \right) > m_\phi/2 \,$ (we do not consider here nonperturbative
decay process, where this bound may not apply). This gives an upper bound
on $\phi_d\,$, which depends on the specific interaction terms between
$\phi$ and $\chi\,$. For interactions of the form $\sigma
\phi\,\chi^2\,+ h.c.$, one finds for the decay rate
\begin{eqnarray}
\Gamma \left(\phi \rightarrow \chi \chi \right) & \simeq & \frac{\sigma^2}{16
\, \pi \, m_\phi^2} \: \sqrt{m_\phi^2 -4 \, m_\chi^2 \left(\phi\right)}
\;\;\;,
\nonumber\\
m_\chi^2(\phi) &=& \sigma \phi \;\;\;\,
\label{decayrate}
\end{eqnarray}
which, equated to the expansion rate of the Universe, gives
\begin{equation}
\frac{\phi_d}{m_\phi} \simeq\frac{m_\phi}{\sigma}\left( -2 r^2+ r \sqrt{1+4 r^2}
\right)\;\;\;,
\label{model1}
\end{equation}
where we have defined
\begin{equation}
r \equiv \sqrt{\frac{3}{8\,\pi}}\,\frac{\sigma^3 M_{Pl}}{16
\pi m_\phi^4}
\sim 0.007 \, \frac{\sigma^3 \, M_{Pl}}{m_\phi^4} \;\;\;.
\end{equation}
We are assuming that the scalar $\phi$ decays after it dominates. Using
Eq.~(\ref{hubble}), this translates in the bound
\begin{equation}
\phi_d < \phi_0 \left( \frac{\phi_0}{M_{Pl}} \right)^3 \;\;\;.
\label{deccond}
\end{equation}
We show in Fig. \ref{livelli} how the ratio $\phi_d/m_\phi$ changes for
different values of $m_\phi$ and $\sigma\,$. As it is clear from the above
discussion, see for example Eq.~(\ref{qincond}), this is a crucial
quantity, since an increase of $\phi_d$ over $m_\phi$ corresponds to an
increase of the number of produced quanta over their energy. We first note
that, for any fixed $m_\phi\,$, there exists only a finite interval for
$\sigma$ where $\phi_d$ is enhanced. This is easily understood, since both
a very low and a very high $\sigma$ reduces the $\phi$ decay rate
(\ref{decayrate}), so that when the decay eventually takes place the $\phi$
amplitude has been strongly damped by the expansion of the Universe. The
decrease of $\Gamma(\phi \rightarrow \chi \chi)$ with decreasing $\sigma$
is rather obvious, since $\sigma$ measures  the strength of the
$\chi$-$\phi$ coupling. In the limit of high $\sigma$, Eq. (\ref{model1})
reduces instead to $\phi_d / m_\phi
\simeq m_\phi/
\left(4 \, \sigma\right)\,$. This shows that, in this limit, the $\phi$
particles are prevented from decaying by the high effective mass $m_\chi
\left( \phi \right)\,$, and only once $\phi_d$ is sufficiently reduced the
decay can occur.

For a given $m_\phi\,$, the maximum of $\phi_d$ corresponds to $r
=1/\sqrt{5}\,$, that is for $\sigma_*$ such that
$\sigma_*/m_\phi \simeq 4 \left( m_\phi / M_{Pl}
\right)^{1/3}\,$. From Fig. \ref{livelli} it is possible to understand
the range for the coupling between the two species such that a sizeable
condensate would appear. If $\sigma$ is not too far from this value, we
have from Eq. (\ref{model1})
\begin{equation}
\frac{\phi_d}{m_\phi} \simeq 0.05 \, \left( \frac{M_{Pl}}{m_\phi} \right)^{1/3}
\;\;\;,
\label{bestcase}
\end{equation}
that is $\phi_d/m_\phi$ increases as $m_\phi$ decreases, as also clear from
Fig. \ref{livelli}. Therefore, formation of a condensate is favoured at small
$m_\phi\,$. For $\sigma$ close to the above value, the necessary condition Eq.
(\ref{cond3}), and the more stringent condition of Eq. (\ref{qincond}) can
indeed be cast in the form
\begin{eqnarray}
f^2 &\gsim& 1.4 \, h \left(\frac{m_\phi}{M_{Pl}} \right)^{1/3}
\;\;{\rm and}\;\; f^2 \gsim (2 \div 5) \left(\frac{m_\phi}{M_{Pl}} \right)^{1/3}
\;\;\;.
\end{eqnarray}
However, both the absolute value of $\phi_d$ and of $Q \simeq f \, m_\phi \,
\phi^2$ decrease as $m_\phi\,$ becomes smaller.  Hence, depending on which
cosmologically scales one is interested in, the mass of the field $\phi$
cannot be taken arbitrarily small, if the condensate is supposed to play any
relevant role. We will discuss this point in more details in the following
section.

Until now we have simply assumed that, when all necessary conditions are
fulfilled, the condensate forms due to a fast thermalization of the $\chi$
quanta. The actual generation of the condensate can be discussed by
studying the typical timescales of self-interactions of the $\chi$
particles.~\footnote{For recent studies in thermalization, see
\cite{terma1,terma2,terma3}.}

Kinetic equilibrium among the quanta of $\chi$ is largely governed by $2
\leftrightarrow 2$ scatterings, allowed at tree level by the interaction term
(\ref{inter}). At the initial stages, when the $\chi$ are produced by the
$\phi$ field the corresponding rate can be estimated to be
\begin{equation}
\Gamma_{2 \leftrightarrow 2} \sim h^2 \; n / \langle E \rangle^2 \sim h^2 \,
\phi_d^2 / m_\phi \;\;\;,
\end{equation}
where $n \geq Q$ is the number density of the particles in the initial
state (initially $n \sim n_\phi$ given in Eq.~(\ref{prochi})) and $\langle
E \rangle$ their average energy, initially of the order of $m_\phi$.
However, a complete thermalization requires that also processes which
violate the particle numbers are in equilibrium. Among these, the ones
which appear at the lowest order in the coupling $h$ are  $4
\leftrightarrow 2$ processes, whose rates can be estimated as follows
\begin{equation}
\Gamma_{4 \rightarrow 2} \sim h^4 \; n^3 / \langle E \rangle^8 \;\;\;,
\Gamma_{2 \rightarrow 4} \sim h^4 \; n / \langle E \rangle^2 \;\;\;.
\end{equation}

For a thermal distribution without a condensate, one has the simple
relations $n \sim \langle E \rangle^3 \sim T^3\,$. However, we will always
consider systems for which $n > \langle E \rangle^3\,$. In the initial
configuration, this occurs for
\begin{equation}
\phi_d > m_\phi \;\;\;.
\end{equation}
Actually, we typically require $\phi_d \gg m_\phi\,$, i.e., as we have seen,
small couplings $\sigma < m_\phi$, and $m_\phi \ll M_{Pl}$, see Eq.
(\ref{model1}). The initial distribution of $\chi
\;,\; {\bar \chi}$ is thus characterized by high particle density and small
energy of the individual quanta, so that the first stages of thermalization
proceed via particle fusion. The decrease of $n$ and increase of $\langle E
\rangle$ have the general effect of decreasing the interaction rates. However,
the condition $n > \langle E \rangle^3\,$ can be never violated if the
system has a sufficiently high charge $Q > \rho^{3/4}\,$. This is
equivalent to require that most of the particle in the final distribution
will be stored in the condensate, Eq.~(\ref{qincond}). When this condition
is met, all the interaction rates are thus always greater than the ones the
$\chi$, $\bar{\chi}$ would have {\it if they were} in a thermal
distribution with the same energy $\rho_\chi \simeq m_\phi^2
\phi_d^2\,$, and without any condensate.
If we therefore evaluate all relevant rates in the latter case,
this will be a sufficient condition to ensure that the system we are
considering rapidly approaches its equilibrium form.

Among the interactions we have considered,  $2 \rightarrow 4$ processes
have the lowest interaction rate. For a thermal distribution with
temperature $T \sim
\left( m_\phi^2 \, \phi_d^2 \right)^{1/4}\,$, we can estimate $\Gamma_{4
\leftrightarrow 2} \sim h^4 \,T\,$, so that thermalization is ensured,
{\it a fortiori}, for
\begin{equation}
h \gsim h_* \equiv \left( \frac{\sqrt{m_\phi \, \phi_d}}{M_{Pl}}
\right)^{1/4}
\;\;\;.
\label{limith}
\end{equation}
If the coupling $h$ satisfies this bound, the $\chi
\;,\; {\bar
\chi}$ system is expected to quickly thermalize right after it is formed at
the decay time $t_d$, leading to the formation of the condensate, mainly
populated by $\chi$ quanta,  and a thermal distribution of both $\chi$ and
$\bar{\chi}$.

For $h \ll h_*$ thermalization can occur only when the scale factor of the
Universe becomes $a \sim a \left( t_d \right) \left( h_* / h \right)^4$,
since the ratio $\Gamma/H$ increases linearly with $a$. Therefore in this
case the final effects of the condensate is expected to be strongly damped
by the dilution of $Q_c$, due to the expansion of the Universe. The case $h
\sim h_*\,$ requires more care,  since (as we remarked) self-interactions
between particles which are forming the condensate are much more efficient
than in the thermal case ($n \gg \langle E
\rangle^3$) and thus the formation of the condensate may be expected to hold
in this case as well. However, we believe that a more precise answer in
this regime cannot be obtained without a numerical study of the relevant
Boltzmann equations. For this reason, hereafter the relation $h \gsim h_*$
will be always assumed.

Before concluding this section, it may be convenient to summarize the above
results. We have considered some light scalars $\chi\,$, ${\bar \chi}$ with
an initial asymmetry $Q\,$. Eq.~(\ref{cond3}) is the necessary condition
for a condensate to appear in the distribution of $\chi\,$. The
distributions for $\chi$ and ${\bar \chi}$ also keep a regular part
(\ref{bosein}), which contributes to the energy density of $\chi$ quanta.
The total energy density redshifts as the one of radiation, as it is clear
from Eqs.~(\ref{rhotot}) and (\ref{truemass}). The
condition~(\ref{qincond}) corresponds to the case where most of the
particle in the final distribution are stored in the condensate, so that
the final number density approximatively coincides with $Q_c\,$. Of course
this condition is a stronger requirement than just the formation of the
condensate of Eq.~(\ref{cond3}). In the specific example we have discussed,
the quanta of $\chi$ are produced by the decay of a massive scalar
$\phi\,$. We parametrized the energy density and the decay time of $\phi$
by its classical amplitude $\phi_d$ when it decays, given by
Eq.~(\ref{model1}). We have assumed that $\phi$ decays after it dominates,
which is ensured by Eq.~(\ref{deccond}). Finally, (\ref{limith}) is a
sufficient condition to guarantee fast self-interactions among the quanta
of $\chi
\,,\, {\bar \chi}\,$, so that the condensate actually forms right after the
decay of $\phi\,$.

\section{Cosmological implications}

The aim of this section is to discuss the possible implications of the
condensate of $\chi$ particles. In particular, we show that it may lead to
symmetry restoration. As it is well known, this issue has crucial
consequences for cosmology, since when symmetry eventually breaks again,
once the expansion of the Universe dilutes the condensate, topological
defects can be formed, reintroducing a monopole problem. The possibility of
symmetry restoration after nonperturbative creation at {\em preheating} has
been emphasized in several
works~\cite{prehrest1,prehrest2,prehrest3,prehrest4,prehrest5,prehrest6,prehrest7}.
Here we show that this may happen also in the ordinary, perturbative {\em
reheating} stage, which is generally expected to occur in all models of
inflation.

We suppose that the field $\chi$ is coupled to some other scalar field
$\psi\,$. For definiteness we will consider $\psi$ to be complex, though
our result applies to a real scalar field as well. Besides this
interaction, the field $\psi$ is assumed to have the potential
\begin{equation}
V_\psi = \kappa \left( |\psi|^2 - \psi_0^2 \right)^2 \;\;,
\end{equation}
and to be already settled down in a state with $|\langle \psi \rangle| =
\psi_0$.

Due to the coupling to the $\psi$ field, the scalar $\chi$ acquires in
general an effective mass $m_\chi\left(\psi\right)$ and a modi--fied
dispersion relation, $\omega^2 = {\bf k}^2 + m_\chi^2(Q_c,T) + m_\chi^2
\left(\psi
\right)$, due to forward-scatterings, which however do not modify the
distribution function of the $\chi$ quanta as long as $m_\chi(Q_c) \gg
m_\chi(\psi)$. On the other hand this coupling also results in a
contribution to the $\psi$ one-loop effective potential
\cite{deltav1,deltav2},
\begin{eqnarray}
\frac{\partial \, \Delta V_\psi}{\partial \, m_\chi^2 \left( \psi \right)}
&=& \frac{1}{\left( 2 \, \pi \right)^3} \int d^4 p \: \delta \left( p^2
-m_\chi^2(Q_c,T)-m_\chi^2 \left( \psi \right) \right) {\cdot} \nonumber \\
&&\left( f_\chi
\left( \bf{p}
\right) + f_{\bar{\chi}} \left( \bf{p}
\right) \right)\;\;.
\end{eqnarray}
In case the $\chi$ number density is dominated by the condensate, $f_\chi(
{\bf{p}}) \sim  (2 \pi)^{3} Q_c \delta^3 \left( {\bf{p}} \right)$, this
relation gives
\begin{equation}
\Delta V_\psi =  \sqrt{m_\chi^2(Q_c,T) + m_\chi^2 ( \psi )
}\, Q_c\,\,\,,
\label{deltav}
\end{equation}
where as discussed before we are considering $Q_c \sim n_\chi \gg
n_{\bar{\chi}}$.

This term may shift the ground state of $\psi$ to zero, thus restoring any
symmetry which may be spontaneously broken when instead $|\langle \psi
\rangle |= \psi_0$.

As an example, we consider an interaction term of the form $g^2
\,|\psi|^2 \, |\chi|^2$. In order to use the analysis contained in the
previous sections, we first have to constrain our study to the case where
the $\chi$ quanta mass is still dominated by the radiative term of Eq.
(\ref{masschi}), i.e.
\begin{equation}
g \psi_0 < (2 h Q_c)^{1/3} \;\;\;.
\label{condition1}
\end{equation}
{} From Eq. (\ref{deltav}) we therefore get
\begin{equation}
\Delta V_\psi \sim  \left(m_\chi(Q_c,T)+\frac{g^2 \psi^2}{2 m_\chi(Q_c,T)} \right)
\,Q_c\,\,\,,
\label{deltavbis}
\end{equation}
which shifts the minimum of the total potential $\,V_\psi +
\Delta V_\psi\,$ to  $\psi = 0$ for
\begin{equation}
g^2 Q_c \, \gsim 4 \kappa \,m_\chi \left( Q_c \right)\, \psi_0^2 \;\; .
\label{rest}
\end{equation}

The charge per unit volume $Q_c$ takes its largest value at the decay of
the $\phi$ field introduced in the previous section, and is then diluted by
the expansion of the Universe. To see whether the symmetry broken by $\psi$
is restored we evaluate the conditions (\ref{condition1}) and (\ref{rest})
at the decay time $t_d$. Using the expression (\ref{prochi}) for $Q_c$, we
have
\begin{equation}
\frac{\psi_0}{\left( m_\phi \, \phi_d^2 \right)^{1/3}} \lsim {\rm min} \; \left[
\frac{\left( 2 \, h \, f \right)^{1/3}}{g} \:,\:   \frac{f^{1/3} \, g}
{2^{7/6}\kappa^{1/2}h^{1/6}}
 \right] \;\;\;.
\label{rest2}
\end{equation}
For definiteness, and to get the largest scale which can be affected by the
formation of the condensate, we consider the case where the ratio
$\phi_d/m_\phi$ is the largest compatible with Eq. (\ref{model1}), for each
$m_\phi$. This is equivalent to the choice $\sigma_*/m_\phi \,$, so that
from Eq. (\ref{bestcase}) we have
\begin{equation}
\frac{\psi_0}{m_\phi} \lsim \left( \frac{M_{Pl}}{m_\phi} \right)^{2/9} \: {\rm
min} \; \left[ 0.2 \frac{\left( h \, f \right)^{1/3}}{g} \:,\: 0.06
\frac{f^{1/3} \, g}{\kappa^{1/2}h^{1/6}}  \right] \;\;\;.
\label{rest3}
\end{equation}

Once some specific values for the parameters are chosen, Eq.~(\ref{rest3})
indicates which is the maximal scale $\psi_0$ which can be restored by the
condensate. It is not unconceivable that scales as high as the GUT one can
be restored, provided $m_\phi$ is taken sufficiently high and the coupling
$\sigma$ is fixed to give the maximal $\phi_d$ as in Eq.~(\ref{bestcase}).
On the contrary, if we are interested in restoring  a symmetry at a smaller
scale, more freedom is allowed for $m_\phi\,$, as well as for the value of
$\phi_d
\,$, as it is clear from Eq.~(\ref{rest2}). We remind that these
conditions must be supplemented by the ones found in the previous section,
essentially Eq.~(\ref{cond3}), necessary condition for the condensate, and
Eq.~(\ref{limith}), implying a rapid thermalization of the system. It is
worth noticing that these constraints are all mutually compatible.

\section{Conclusions}

In this paper we have considered the possibility that a light scalar field,
whose quanta are excited at a perturbative reheating stage, may produce a
large Bose-Einstein condensate. This phenomenon takes place if these quanta
carry a conserved charge and the reheating stage produces large value for
this charge. The motivation for this study relies on the fact that in
scenarios as the Affleck-Dine models for baryon an lepton number
generation, the decay of a modulus field may typically result in a quite
large charge per comoving volume in the final relativistic degrees of
freedom, and this charge is conserved or only weakly violated by the
self-interaction processes after the reheating stage.

We have studied in details the conditions which should be fulfilled for the
formation of the condensate for a complex scalar field with a simple
quartic self-interaction term, and how the condensation process depends on
the time scale for thermalization via scattering and number changing
processes. For a wide range for the mass of the field whose decays lead to
reheating and for the relevant coupling constants, the condensation
phenomenon may easily occur and, moreover, the condensed particles can
represent the main contribution to the total number per comoving volume.

We have finally considered the role that the condensate may have in restoring
a symmetry which was broken at a larger energy scale. It is interesting to
notice that also a grandunified symmetry, with scale $M_{GUT} \sim 10^{15}
\div 10^{16}~GeV$, can in principle be restored, although this requires some
fine tuning in the parameters of the model. Restoration can occur more
naturally for smaller scales, until the expansion of the Universe dilutes
the charge per comoving volume which is stored in the condensate. This
possibility would represent a new intriguing scenario in the evolution of
the primordial Universe.

\acknowledgments
We are pleased to thank A. Riotto and R. Allahverdi for interesting
conversations and valuable comments. The work of M.P. is supported by the
European Commission RTN programme HPRN-CT-2000-00152. S.P. is supported by
the European Commission by a {\it Marie Curie} fellowship under the
contract HPMFCT-2000-00445.

\end{document}